\documentclass[aps, reprint, nofootinbib]{revtex4-1}
\usepackage[sort&compress]{natbib}
\usepackage{amsmath}
\usepackage{graphicx}
\usepackage[rightcaption]{sidecap}
\usepackage{courier}
\usepackage{hyperref}

\pdfoutput=1
\begin{document}

\title{A Holographic Dual of the Quantum Inequalities}
\author{Adam Levine}
\email[]{arlevine@berkeley.edu}
\affiliation{Department of Physics, University of California, Berkeley, CA 94720, USA}

\begin{abstract}
In this note, we establish the 2-D Quantum Inequalities - first proved by Flanagan - for all CFTs with a causal holographic dual. Following the treatment of Kelly \& Wall, we establish that the Boundary Causality Condition in an asymptotic AdS spacetime implies the Quantum Inequalities on the boundary. Our results extend easily to curved spacetime and are stable under deformations of the CFT by relevant operators. We discuss higher dimensional generalizations and possible connections to recent bounds on $a/c$ in 4-D CFTs.
\end{abstract}

\maketitle

\section{Introduction and Summary}
In classical general relativity, energy conditions prove essential in determining causal structure of spacetime \cite{Hawking and Ellis}. Penrose used energy conditions to prove key singularity theorems \cite{Penrose}. Prevalent examples of energy conditions include the null energy condition (NEC) and the averaged null energy condition (ANEC). The NEC bounds the null-null component of the stress tensor at each point $T_{kk} \geq 0$ and the ANEC bounds the integral over a null line
\begin{equation}
\int{T_{kk}\,d\lambda}\geq 0
\end{equation}
with $\lambda$ an affine parameter.

However, in quantum mechanics these energy conditions no longer hold as fluctuations can contribute to negative energy density. For general states in quantum field theory, the expectation value of the stress tensor at a point $T_{kk}$ is unbounded below. Certain quantum extensions of the NEC have been proposed \cite{QNEC, holoQNEC}, demonstrating a deep connection to the entanglement structure of quantum states. In \cite{Roman, Roman2, QIC}, ANEC-type inequalities were initially proposed to hold for two and four-dimensional, free, massless quantized scalar fields. Flanagan improved these two-dimensional Quantum Inequalities (QIs) by more general means \cite{Flanagan}. This result, which we will  prove in a holographic context, states that 
\begin{equation}\label{eq:QIreal}
\int_{-\infty}^{\infty}{\rho(u) \langle T_{uu}\rangle\,du} \geq -\frac{1}{48\pi}\int_{-\infty}^{\infty} {\frac{(\rho^{\prime}(u))^2}{\rho(u)}\,du}
\end{equation}
where $\rho$ is any smooth, positive smearing function which can have compact support. In regions where $\rho(u)=0$, $\rho^{\prime}(u)^2/\rho(u)$ is set to zero.

Energy conditions have played an important role in illuminating gauge/gravity duality \cite{AdS/CFT}. The quantum null energy condition was proven within the context of holography as originating from a property of Ryu-Takayanagi surfaces \cite{holoQNEC}. The quantum generalization of the ANEC was also investigated by Kelly \& Wall in \cite{WallandKelly}. They proved holographically that the ANEC for CFTs living on the boundary of asymptotic AdS follows from the requirement of ``good" causal properties of the duality: signals on the boundary of AdS cannot propagate faster than the speed of light by using the bulk as a shortcut. 

This property was first established in \cite{Gao}, in a theorem we will refer to as the Gao-Wald theorem. The theorem was proven for a generic spacetime with a time-like, conformal boundary that satisfies the null curvature condition. One can prove that this boundary causality condition (BCC) implies the Quantum Inequalities as well as the ANEC in the boundary, where we conform with the terminology found in \cite{Engelhardt}. 

The proof laid out here will follow the derivation in \cite{WallandKelly}. We begin by reviewing the proof of the two-dimensional Quantum Inequalities for a general CFT. We then show how the bulk-boundary causality condition implies this inequality in the holographic context. We conclude with speculation about higher dimensional generalizations of this proof as well as a curved spacetime analog.

\section{Proof of Quantum Inequalities for a 1+1-D CFT}
In this section, we review a field-theoretic proof of the two dimensional Quantum Inequalities. Flanagan proved the Quantum Inequalities for massless and massive, free scalar field theory \cite{Flanagan}. This proof was later extended to all two-dimensional CFTs in \cite{Hollands}. We begin by recalling the transformation properties of the stress-energy tensor under conformal transformations, which will prove crucial in understanding the nature of these inequalities. 

Formally, a conformal transformation is a map between two Riemannian manifolds, $\phi: (M,g) \to (\hat{M},\hat{g})$ such that $\phi^*\hat{g}=e^{2\sigma} g$. In flat space, we can write the metric as $ds^2 = -dUdv-dvdU$. If we make the coordinate transformation $u \mapsto U=U(u)$, flat space is conformally related to itself: 
\begin{equation}
d\,\hat{s}^2 = U^{\prime}(u)(-du\,dv-dv\,du)=U^{\prime}(u)ds^2
\end{equation}
Schematically then, our Weyl transformation takes $$ U^{\prime}(u)(-du\,dv - dv\,du) \mapsto -du\,dv-dv\,du$$ so that $\sigma = \frac{-1}{2} \log(U^{\prime})$. 

By the work of \cite{Higherd}, we know that correlation functions involving the stress tensor $T_{uu}$ on the source manifold are related to correlation functions involving the stress tensor on the target space as so :
\begin{equation}\label{eq:conformal}
\langle T_{uu}(u)\Phi(u) \rangle_U = \langle T_{uu} \Phi(u) \rangle_u + \frac{c}{24\pi}(\sigma_{uu} +\sigma_u ^2)\langle \Phi(u) \rangle_u
\end{equation}
where we have defined: $\sigma_{\alpha} \equiv \partial_{\alpha} \sigma$ and $\sigma_{\alpha\beta} \equiv \partial_{\alpha} \partial_{\beta} \sigma$. Here the state subscripts denote whether we are taking vacuum expectation values with respect to the $U$-coordinate vacuum or the $u$-coordinate vacuum. \footnote{Cappelli \& Coste actually did not prove this result. They showed that the one-point function of the stress tensors should be related in this way. We believe it to be true that their method of proof extends easily to correlation functions with general field insertions.}

For the remainder of the discussion, we assume that $U(u)$ is bijective and smooth and so defines an endomorphism, $S$, that takes one vacua to the other: $$S|u\rangle = |U\rangle$$ There are subtleties related to the existence of this operator for general functions $U(u)$, but we let the interested reader consult \cite{Flanagan} for details. Meantime, we only utilize functions, $U$, that satisfy the necessary conditions laid out therein.

After introducing this $S$ map, we can derive an operator equation from (\ref{eq:conformal}) that states:
\begin{equation}\label{eq:operator}
S^{\dag}T_{uu}(u)S = \hat{T}_{uu}(u) + \frac{c}{12\pi}(\sigma_{uu} +\sigma_u ^2)
\end{equation}
where the hat signifies that this stress tensor lives on the target manifold. 

Conformal invariance of the theory implies that the right hand side transforms covariantly: $\hat{T}_{uu}(u)=\hat{T}_{UU}(U)\,U^{\prime}(u)^2$ as an operator. In order to get the quantum inequalities, we must integrate \ref{eq:operator} against a given sampling function, $\rho(u)$. If one plugs this equation back in to \ref{eq:operator} and integrates we find an equation first proved in \cite{Flanagan}:
\begin{multline}\label{eq:final}
\int_{-\infty}^{\infty} \rho(u) S^{\dag}T_{uu}(u)S\,du=\int_{-\infty}^{\infty} \rho(u)\hat{T}_{UU}(U)U^{\prime}(u)^2\,du\\ 
+\frac{c}{24\pi}\int_{-\infty}^{\infty} \rho (u) (\sigma_{uu} +\sigma_u ^2)\,du
\end{multline}
Choosing our coordinate transformation such that $\rho(u)=\frac{1}{U^{\prime}}$ then $\sigma=\frac{1}{2}\log(\rho(u))$. We see that (\ref{eq:final}) just reduces to 
\begin{multline}\label{eq:final2}
\int_{-\infty}^{\infty} \rho(u) S^{\dag}T_{uu}(u)S\,du=\int_{-\infty}^{\infty} \hat{T}_{UU}(U)dU\\ 
-\frac{c}{48\pi}\int_{-\infty}^{\infty} \frac{\rho^{\prime}(u)^2}{\rho(u)} \,du
\end{multline}
As Flanagan notes, the first term on the right hand side of this equation is just $H_R$, the Hamiltonian associated with one chiral sector of the CFT. This is a manifestly positive definite operator for all $U(u)$, and so, importing the fact that $H_R \geq 0$, we establish the Quantum Inequalities found in (\ref{eq:QIreal}). We note that this also happens to be equivalent to the ANEC in two dimensions.  This equivalence in conjunction with the work of Kelly \& Wall leads us naturally to the following section.

\section{Proof of the Quantum Inequalities from Bulk-Boundary Causality}
Understanding the relationship between bulk-boundary causality and energy conditions can shed light on the nature of AdS/CFT. In this note, we take ``good" bulk-boundary causality to have the same definition as in \cite{WallandKelly, Engelhardt} - no curve through the bulk can beat a null curve on the boundary. 

Given that good bulk-to-boundary causality proves the ANEC \cite{WallandKelly}, one should not be surprised at all that it also proves the Quantum Inequalities, but it is nice to see how a reasonable bulk condition explicitly translates to a proven fact in the boundary. Our result generalizes Kelly \& Wall's proof to include a larger class of bulk curves as will become evident. Although the computation of the previous section relied on the transformation properties of the stress-energy tensor, this section uses the machinery of holographic renormalization \cite{Skenderis} to avoid the discussion of conformal transformations.

The proof will follow closely along the lines of \cite{WallandKelly}. We take the AdS metric to have the form 
\begin{equation}\label{eq:metric}
ds^2 = \frac{dz^2 - 2dudv+z^2 \gamma_{ab} dx^a dx^b}{z^2}
\end{equation}
where here $\gamma_{ab} = t_{ab} + z^2 s_{ab}$ with latin indices denoting boundary coordinates. The holographic dictionary \cite{Skenderis} tells us that $t_{ab} \sim \langle T_{ab} \rangle/c$, where $c$ is the central charge.

The set up, as depicted in Fig. 1, will be to pick a null curve, $v(u)=\,$constant with $u\in [-L,L]$, lying at the $z=0$ plane in Poincare coordinates. If an arbitrary time-like or null curve leaves the boundary at the same initial point and dips into the bulk, the BCC says that this curve should be delayed upon arriving at the boundary. Indeed, the change in the null coordinate orthogonal to $u$ obeys 
\begin{equation}\label{eq:BCC}
\Delta v\geq 0
\end{equation}
for any time-like or null curve starting on the boundary. We now show that this inequality implies the quantum inequalities.

We can parameterize our bulk curve also in terms of $u$, and so we pick a curve dipping into the bulk such that 
\begin{multline}\label{eq:curve}
z(u) = \epsilon \sqrt{\rho (u)},\ 
V(u) = \frac{\epsilon^2}{2}  \int_{-L}^{u} \rho(u') \gamma_{uu}(u')\,du' \\ +\frac{\epsilon^2(1+\delta)}{8} \int_{-L}^{u} \frac{(\rho^{\prime}(u'))^2}{\rho(u')} \,du'
\end{multline}
where we have introduced the small parameters $\delta,\,\epsilon$. For $\delta=0$, the curve is null at leading order and so higher order terms become important. To assure that (\ref{eq:curve}) is timelike, we pick $\delta>0$ but eventually take the limit as $\delta \to 0$.

\begin{figure}[]\label{fig:curve}
	\includegraphics[width=.35\textwidth]{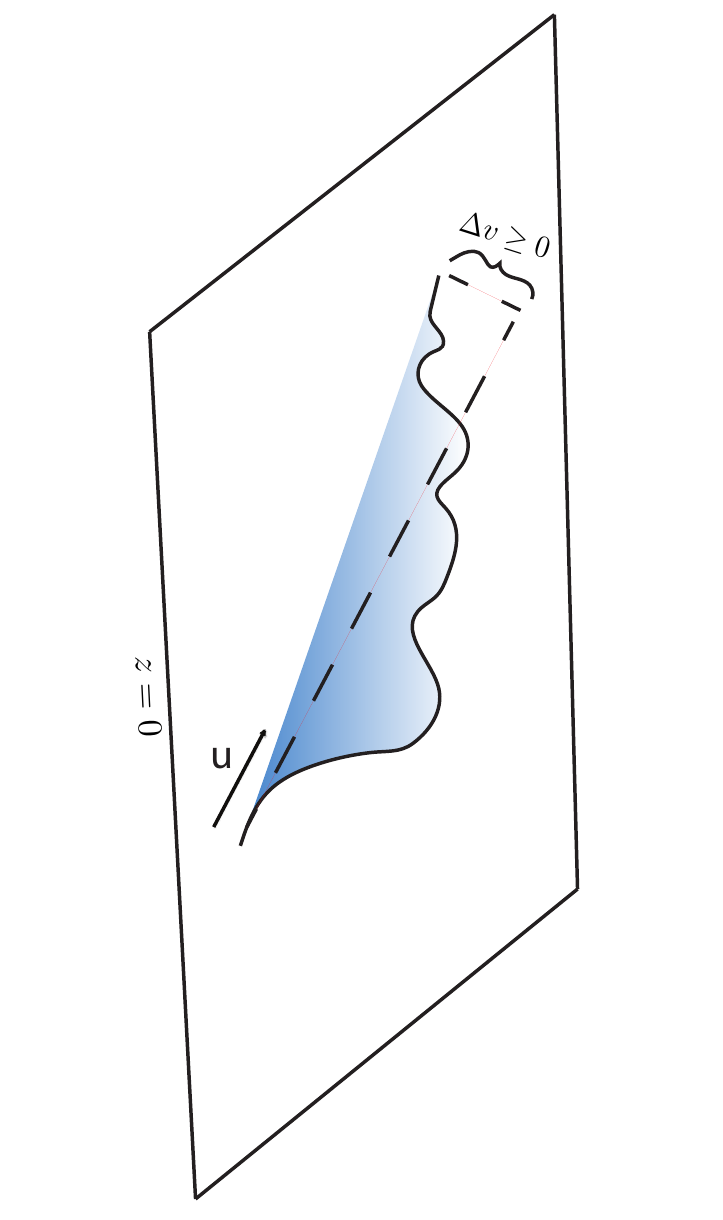}
	\caption{A sample curve with bulk profile given by $z(u)=\epsilon \sqrt{\rho(u)}$ for $u\in [-L,L]$. The curve originates at the point $(u,v,z)=(-L,0,0)$ and is constrained to land inside the future light cone of the point $(u,v,z)=(L,0,0)$.}
\end{figure}
Expanding in $\epsilon$, this curve is manifestly timelike at leading order
\begin{multline}\label{eq:timelike}
Z^{\prime}(u)^2 -2V^{\prime}(u) + Z^2(u) \gamma_{uu} (u) + \mathcal{O}(\epsilon^3)=\\
\frac{-\epsilon^2\delta}{4} \frac{\rho^{\prime}(u)^2}{\rho(u)} + \mathcal{O}(\epsilon^3)
\end{multline}
Thus, one can apply the BCC to this curve; this curve must be delayed by some non-negative amount with respect to its boundary competitor.
Sending $\epsilon \to 0$ and imposing the condition in \ref{eq:BCC}, one finds
\begin{equation}\label{eq: QI}
\int_{-L}^L\rho(u^{\prime})\gamma_{uu}\,du^{\prime} \geq -\frac{1+\delta}{4}\int_{-L}^{L} \frac{\rho^{\prime}(u^{\prime})^2}{\rho(u^{\prime})}\,du^{\prime}
\end{equation}

The left hand side can be brought into a familiar form by quoting the standard holographic result \cite{Skenderis} $$\gamma_{uu} = \frac{12\pi}{c}\langle T_{uu} \rangle$$ for $d=2$. Plugging this in, we get the result that
\begin{multline}\label{eq:QIfinal}
\int_{-L}^L \rho(u^{\prime})\langle T_{uu} (u^{\prime}) \rangle\,du^{\prime} \geq -\frac{c(1+\delta)}{48\pi}\int_{-L}^L \frac{\rho^{\prime}(u^{\prime})^2}{\rho(u^{\prime})}\,du^{\prime}
\end{multline}
Since this must hold for all $\delta$ and $L$, we can with impunity take the $L\to \infty$ and $\delta \to 0$ limits to recover (\ref{eq:QIreal}).

A few comments are in order. Throughout this discussion, we have assumed that $\rho(u) \geq 0$, and thus $U^{\prime}(u)$ is non-negative everywhere. We have also implicitly assumed that $\rho(u)$ vanishes in the far past and future. Furthermore, whenever $\rho(u)=0$, then we set $\rho^{\prime}(u)^2/\rho(u)$ to zero by hand. This is a reasonable assumption for most smooth sampling functions of interest. From a bulk point of view, this just means that the curve smoothly lands on the boundary.

This entire analysis is stable under relevant deformations of the CFT as their effects show up at higher order in the perturbative Fefferman-Graham expansion. Furthermore, the extension to a curved background on the boundary should be straightforward and the result of \cite{Flanagan2} should be exactly recovered. To see this, we note that for a curved background the Fefferman-Graham expansion becomes \cite{Skenderis}
\begin{equation}\label{eq:feff}
\gamma_{uu} = \frac{1}{2}\left ( R g_{uu} + t_{uu} \right)
\end{equation}
where $t_{uu}$ is defined as before. If the curve on the boundary is null with respect to the boundary metric, $g_{uu} = g_{\mu\nu}u^{\mu}u^{\nu}$ will vanish. Thus, the Ricci scalar term disappears and the proof goes forward as before. This matches with the result of \cite{Flanagan2} up to the factor of the central charge.

We can also turn this inequality around. By taking the limit of Gaussians, we can arrive at a pointwise energy condition. In general, the lower bound with till tend to $-\infty$ for $c>0$. If, however, $c<0$, then we get that the stress tensor blows up at every point along this null line. Thus, if we take the Quantum Inequalities to be true, then $c<0$ is seen as unphysical. Note that the CFT proof required no assumptions on $c$. We know also that $c>0$ follows from unitarity of the CFT and so the Quantum Inequalities appear to make contact with this constraint. In higher dimensions, this discussion may bear more insight as discussed below.  

\section{Discussion}
In this section, we discuss the potential generalization to higher dimensions. Immediately one runs into power-counting issues if the above approach is naively applied to four dimensions. Of course, this is no surprise as the integral along a curve of the stress tensor has no lower bound in higher- dimensional theories \cite{Wall}. Quantum inequalities analogous to those in 2-dimensions have been formulated in 3+1-D for massive and massless scalar field theory \cite{4DQI}, but these inequalities involve the time-like component of the stress-tensor - as opposed to the null component. Fewster \& Eveson found that the stress tensors for these theories obey
\begin{equation}\label{eq:4D}
\int \rho(t)T_{tt}(t) \,dt \geq -\frac{1}{16\pi^2} \int (\rho(t)^{1/2\,\prime \prime})^2\,dt
\end{equation}
Although the 2-D QIs are optimal, these 4-D inequalities may not be, but we can still try to follow the same logic of section 2, inspired by \cite{Flanagan}. In four-dimensions, the stress tensor transforms analogously to the lower dimensional case \cite{Higherd}. If we consider the conformal transformation
\begin{multline}\label{eq:4dtransform}
-dT^2 + d\vec{x}^2 = -T^{\prime}(t)^2\,dt^2 + d\vec{x}^2 \mapsto
-dt^2 + \frac{d\vec{x}^2}{T^{\prime}(t)^2}
\end{multline}
so that $\sigma(t) = -\log(T^{\prime})$. Then using equation (4.24) from \cite{Higherd}, we see straightforwardly that 
\begin{multline}\label{eq:QI4}
\frac{\langle \hat{T}_{tt}(t) \rangle}{T^{\prime}} - T^{\prime}\langle T_{tt}(t) \rangle = \\ 
(T^{\prime})^2 \left [ \frac{3a}{8\pi^2} \sigma_{tt}\sigma_t^2 + \frac{c}{8\pi^2}\left(\frac{\sigma^{(4)}}{2} - 3 \sigma_{tt}\sigma_t^2\right) \right]
\end{multline}
Applying the coordinate transformation to $T_{tt}$ and integrating against $\rho(t)$, we find that 
\begin{multline}\label{eq:4dQI}
\int \rho(T) \langle T_{TT}(T) \rangle \, dT = \\ \int \rho(t) \frac{\langle \hat{T}_{TT}(T) \rangle}{T^{\prime}}\, dt - \int \rho(t) \Delta(t)\,dt
\end{multline}
where $\Delta(t)$ is defined to be the factor multiplying $T^{\prime}(t)^2$ in (\ref{eq:QI4}). This equation can be interpreted as an operator statement and so the first term on the right hand side does not vanish in general. 

Importantly, however, we do not currently have the ability to bound the right hand side. If an AWEC-type inequality holds for this conformally flat, FRW-type spacetime in (\ref{eq:4dtransform}), then we expect $\rho(t)=T^{\prime}(t)^n$ for some $n$. Plugging this in to (\ref{eq:4dQI}) and integrating by parts, we find
\begin{multline}\label{eq:fourdconformal}
\int \rho(t)\Delta(t)\, dt = \\ \frac{-2(a-c+\frac{n^2}{3}c)}{16\pi^2n^3} \int \frac{(\rho^{\prime})^4}{\rho^3}\,dt + \frac{c}{16\pi^2 n} \int \frac{(\rho^{\prime \prime})^2}{\rho}\,dt 
\end{multline}
where $a,c$ are the central charges of the 4-d theory. Interestingly, if we take $n=4$, one finds that (\ref{eq:fourdconformal}) becomes
\begin{multline}\label{eq:a/c}
\int \rho(t)\Delta(t)\,dt = 
\\\int \frac{c}{16\pi^2} ((\rho^{1/2})^{\prime \prime})^2\, dt - \left( \frac{a+c}{16\pi^2\,32}\right) \int \frac{(\rho^{\prime})^4}{\rho^3}\,dt
\end{multline}
This equation should be compared with (\ref{eq:4D}), particularly in the case of a massless scalar field theory with $c=1$ and $a/c=1/3$. 

We suspect that whatever inequality is discovered for $\int \rho(t) \frac{\langle \hat{T}_{TT}(T)\rangle}{T^{\prime}}\,dt $ will lead to a bound on $a/c$ by an argument analogous to the one given above. This result would be reminiscent of those found in \cite{hofman} and recently proven rigorously in \cite{poland}. A connection between bounds on $a/c$ and spacetime averaged inequalities was also discussed in \cite{Farnsworth}, but we leave a more thorough examination of this point for later work.
\vspace{-1mm}
\section{Conclusion}
We have shown a connection between the quantum inequalities in 1+1-d CFTs and the BCC in the bulk. We have also hinted at a potential connection to the bounds of \cite{hofman} and \cite{poland}. A higher dimensional generalization of the bulk interpretation of our discussion is still lacking. While null energy conditions have been discussed thoroughly in the context of bulk-boundary physics, averaged timelike inequalities remain more mysterious from the point of view of the bulk.
We would also like to better understand the bulk origins of both the Quantum Null Energy Condition, the ANEC  and the Quantum Inequalities. The recent work of \cite{Engelhardt} may help to shed light on this. We hope to further the investigation of all these points in the future.
\vspace{-1mm}
\section*{A\lowercase{cknowledgements}}
We thank C. Akers, R. Bousso, Z. Fisher, E. Flanagan, I. Halpern, W. Kelly, J. Koeller, S. Leichenauer, A. Moghaddam and A. Wall for insightful discussions. The work of AL is supported in part by the Berkeley Center for Theoretical Physics, by the National Science Foundation (award numbers 1214644, 1316783, and 1521446), and by the US Department of Energy under Contract DEAC02-05CH11231.

\bibliographystyle{apsrev4-1} 

\end{document}